\documentclass{nature}
\usepackage{epsfig}
\usepackage{amsmath}
\usepackage{amsmath,mathrsfs,amsbsy,color,graphicx,bm,amsthm,amsfonts}
\usepackage{units}
\usepackage{bbm}
\usepackage{times}
\usepackage{dcolumn}
\usepackage{mathrsfs}
\usepackage{amsmath,amssymb,epsfig,float}

\begin{document}

\title{Photon Devil's staircase: photon long-range repulsive interaction in
lattices of coupled resonators with Rydberg atoms}
\author{Yuanwei~Zhang$^{1,2}$, Jingtao Fan$^{1}$, J.-Q.~Liang$^{2} $, Jie Ma$%
^{1}$, Gang~Chen$^{1,*}$, Suotang~Jia$^{1}$, Franco~Nori$^{3,4}$ }
\maketitle

\baselineskip=0.75 cm
\begin{affiliations}
\item
State Key Laboratory of Quantum Optics and Quantum Optics Devices, Institute of Laser spectroscopy, Shanxi University, Taiyuan 030006, P. R. China \\
\item
Institute of Theoretical Physics, Shanxi University, Taiyuan 030006, P. R.
China \\
\item
Center for Emergent Matter Science, RIKEN, Wako-shi, Saitama 351-0198, Japan\\
\item
Physics Department, University of Michigan, Ann Arbor, Michigan 48109-1040,
USA\newline
$^*$Corresponding author, e-mail: chengang971@163.com\newline
\end{affiliations}

\begin{abstract}
\baselineskip=0.8 cm The realization of strong coherent interactions between
individual photons is a long-standing goal in science and engineering. In
this report, based on recent experimental setups, we derive a strong photon
long-range repulsive interaction, by controlling the van der Waals repulsive
force between Cesium Rydberg atoms located inside different cavities in
extended Jaynes-Cummings-Hubbard lattices. We also find novel quantum phases
induced by this photon long-range repulsive interaction. For example,
without photon hopping, a photon Devil's staircase, induced by the breaking
of long-range translation symmetry, can emerge. If photon hopping occurs, we
predict a photon-floating solid phase, due to the motion of particle- and
hole-like defects. More importantly, for a large chemical potential in the
resonant case, the photon hopping can be frozen even if the hopping term
exists. We call this new phase the photon-frozen solid phase. In
experiments, these predicted phases could be detected by measuring the
number of polaritons via resonance fluorescence. \newline
\end{abstract}

Strong interactions between individual photons play an essential role in
achieving photon quantum information processing\cite{TEN14, DLM04, KA08,
LZC13} as well as in exploring exotic many-body phenomena of light\cite%
{IC13,BI09,IMG14}. In contrast to electrons, interacting directly via
Coulomb repulsion, the photon-photon interactions must be mediated by matter%
\cite{DEC14}. Being an important challenge, the realization of such
matter-mediated interactions has become a long-standing goal in science and
engineering. During the past decades, much theoretical\cite%
{HSE99,GAJ11,SHF90,SHI96} and experiental\cite{TGA14,OFT13} effort has been
made to enhance the nonlinear interaction to a strong regime at the
single-photon level. Moreover, photon-photon interactions can lead to an
on-site photon-blockade effect\cite{IAH97,KMB05}, when each cavity mode
interacts with a two-level atom. By further considering the novel
competition between the on-site photon-blockade effect and the photon
hopping in an array of coupled cavities\cite{MJH08}, quantum simulations\cite%
{BI09,IMG14}, based on the Jaynes-Cummings-Hubbard model\cite{MJH08}, have
studied complex many-body phenomena in condensed-matter and atomic physics,
such as the superfluid-Mott-insulator transition\cite%
{ADG06,MJH06,MJH07,DGA07}, quantum magnetic dynamics\cite{AC07}, glassy
phases\cite{DR07}, solid\cite{JJ13,JJ14} and supersolid\cite{BB13} phases,
and the fractional quantum Hall effect\cite{ALCH12,MH13}.

In this report, based on recent experimental setups, we derive a strong
photon long-range repulsive interaction (PLRRI) by controlling the van der
Waals force between Rydberg atoms located inside different cavities in
extended Jaynes-Cummings-Hubbard lattices. We also find novel quantum phases
induced by this PLRRI. For example, without photon hopping, the breaking of
long-range translation symmetry induces a complex solid structure, i.e., a
photon Devil's staircase. In a \textquotedblleft Devil's staircase", any two
different rational states are separated by many states. If photon hopping
exists, we predict a photon-floating solid phase, due to the motion of
particle- and hole-like defects. More importantly, for a large chemical
potential in the resonant case, photon hopping can be frozen even if the
hopping term exists. We denote this new phase the photon-frozen solid phase.
In experiments, these predicted phases could be detected by measuring the
number of polaritons via resonance fluorescence\cite{KT13}.\newline

\section*{{\protect\LARGE \textbf{Results}}}

\subsection{Extended Jaynes-Cummings-Hubbard model.}

\begin{figure}[t]
\centering\includegraphics[width = 0.7\linewidth]{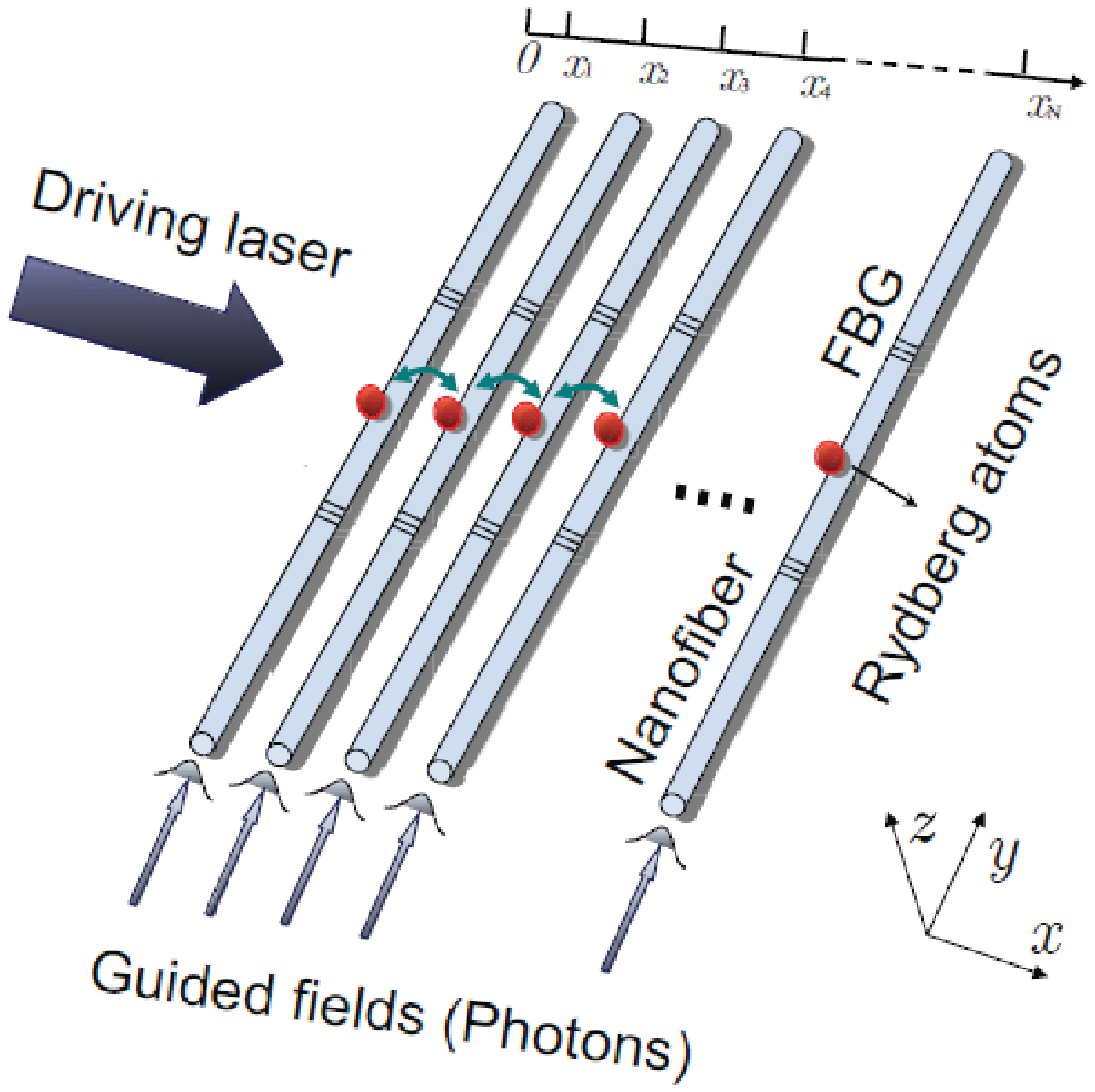}
\caption{\textbf{Schematic diagram of the system studied.}}
\label{fig1}
\end{figure}

\begin{figure}[t]
\centering\includegraphics[width = 1.0\linewidth]{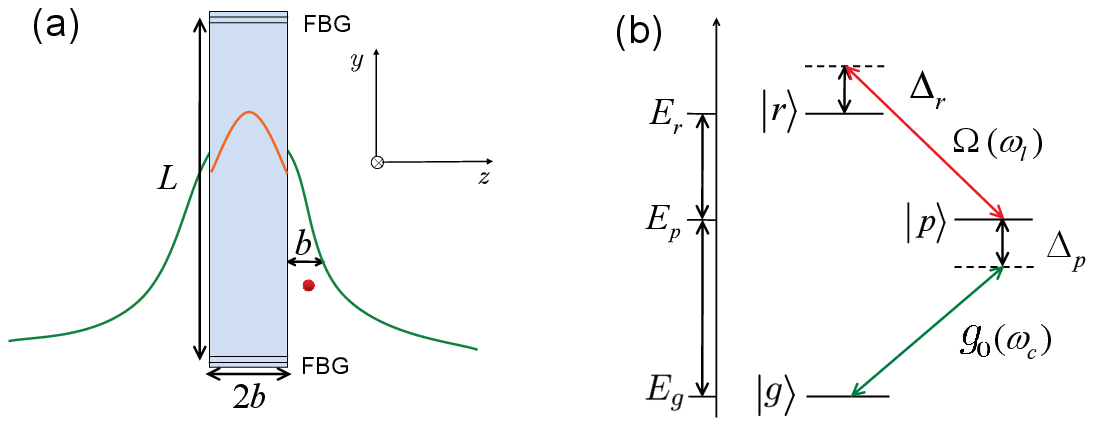}
\caption{\textbf{(a) The sectional plot of the \emph{i}th atom-cavity
interaction system, and (b) energy levels of a single three-level Cs Rydberg
atom and their transition.}}
\label{fig2}
\end{figure}

We first propose a possible way to realize an extended
Jaynes-Cummings-Hubbard model with long-range atom-atom interactions in
different cavities, based on recent experimental setups\cite%
{Goban,Nayak13,Yalla,Vetsch,Nayak14}. As shown in Fig.~\ref{fig1}, a series
of SiO$_{2}$ nanofibers are arranged in the same direction of a specific
plane, and an ensemble of Cesium (Cs) Rydberg atoms are trapped close to
each nanofiber. Each nanofiber, with radius $b=0.25$ $\mu $m, acts as a 1D
photonic crystal cavity, due to its fabricated fiber Bragg-grating (FBG)
structure\cite{Nayak13,Yalla} [see Fig.~\ref{fig2}(a)]. A guided field,
whose evanescent field acts as the quantum cavity mode, propagates along the
cavity $y$ axis. The cavity decay rate is characterized by the parameter $%
\kappa $, which induces the photon hopping in the cavity array\cite{Notomi},
and the distance between nearest-neighbor cavities is about $%
x_{i+1}-x_{i}\approx 2.4$\ $\mu $m. Since the evanescent field strength is
sufficiently weak at the radial distance of about $b$--$4b$ away from the
surface of the nanofiber\cite{Fam05,Fam04}, each adjacent nanofiber pairs
located at such a distance will not lead to an efficient overlap of
different cavity modes, which guarantees that the $i$th ensemble of Cs
Rydberg atoms can interact only with the $i$th cavity\cite{Fam05,Vetsch}.

By using the red- and blue-detuned evanescent light fields around the
optical nanofiber, a two-color optical dipole trap can be formed. This
optical dipole trap should allow an ensemble of Cs Rydberg atoms to be
prepared at a few hundred nanometers from the nanofiber surface\cite%
{Kien,Goban}. For Cs Rydberg atoms, we can choose the fine-structure states $%
\left\vert 6S_{1/2},F=4\right\rangle $ and $\left\vert 6P_{3/2},F^{^{\prime
}}=5\right\rangle $ as the ground state $\left\vert g\right\rangle $ and the
intermediate state $\left\vert p\right\rangle $, respectively, while the
Rydberg state is assumed as $70S_{1/2}$. As shown in Fig.~\ref{fig2}(b), the
photon induced by the evanescent field, with wavelength $852$ nm, governs
the transition between the ground state $\left\vert g\right\rangle $ and the
intermediate state $\left\vert p\right\rangle $, whereas the other
transition between the intermediate state $\left\vert p\right\rangle $ and
the Rydberg state $\left\vert r\right\rangle $ is controlled by a classical
driving laser, with wavelength $510$ nm, as shown in Fig.~\ref{fig1}.

Formally, the total Hamiltonian of the system considered in Fig.~\ref{fig1}
is
\begin{equation}
H=H_{\text{JC}}+H_{\text{HOP}}+H_{\text{V}}-\mu N.  \label{H}
\end{equation}%
In the Hamiltonian (\ref{H}), $H_{\text{JC}}$ describes the interaction
between the photons and the ensemble of Cs Rydberg atoms for all nanofiber
photonic crystal cavities. We first consider the interaction between the
photon and a single three-level Cs Rydberg atom at one cavity. In the
current experimental setups\cite{Goban,Nayak13,Yalla,Vetsch,Nayak14}, the
interaction between photons and the single Cs Rydberg atom is of the order
of MHz (the detailed estimation will be shown in the next subsection).
Therefore, in the framework of the rotating-wave approximation, the
corresponding Hamiltonian is
\begin{equation}
H_{1}=E_{p}\left\vert p\right\rangle \left\langle p\right\vert
+E_{r}\left\vert r\right\rangle \left\langle r\right\vert +g_{0}(a^{\dag
}\left\vert g\right\rangle \left\langle p\right\vert +\text{H.c.})+\omega
_{c}a^{\dag }a+\left[ \Omega \exp \left( -i\omega _{l}t\right) \left\vert
r\right\rangle \left\langle p\right\vert +\text{H.c}.\right] ,  \label{H1}
\end{equation}%
where $E_{p}$ and $E_{r}$ are the energies of the intermediate state $%
\left\vert p\right\rangle $ and the Rydberg state $\left\vert r\right\rangle
$, respectively, $a^{\dag }$ and $a$ are the creation and annihilation
operators of photons with frequency $\omega _{c}$, while $\Omega $ and $%
\omega _{l}$ are the Rabi and driving frequencies of the classical laser,
respectively. When the detuning is large, we can adiabatically eliminate the
intermediate state $\left\vert p\right\rangle $, and rewrite the Hamiltonian
(\ref{H1}) via a unitary transformation as%
\begin{equation}
H_{2}=\omega a^{\dag }a+\epsilon \left\vert r\right\rangle \left\langle
r\right\vert +g_{1}(a^{\dag }\left\vert g\right\rangle \left\langle
r\right\vert +\text{H.c.})+\lambda a^{\dag }a\left\vert g\right\rangle
\left\langle g\right\vert ,  \label{H2}
\end{equation}%
where $\omega =\omega _{c}-\omega _{l}$ is the effective photon frequency, $%
\epsilon =E_{r}-E_{g}-\omega _{l}+\Omega ^{2}/\Delta _{p}$ is the effective
transition frequency of the two-level Rydberg atom, $g_{1}=g_{0}\Omega
/\Delta _{p}$ is the effective interaction strength, and $\lambda =$ $%
g_{0}^{2}/\Delta _{p}$. For large detuning, $\lambda $ is very small and we
thus can omit the interaction term $a^{\dag }a\left\vert g\right\rangle
\left\langle g\right\vert $.

In addition, for large detuning, $g_{1}$ is also weak. In order to enhance
the effective atom-photon interaction strength, here we consider an ensemble
of Cs Rydberg atoms in the center of each cavity. For simplicity, we also
assume that the number of Cs Rydberg atoms in each cavity is a constant $%
N_{R}$. The strong van der Waals repulsive interaction between Cs Rydberg
atoms in the same cavity generates a Rydberg-blocked effect, which excites
only one Cs Rydberg atom\cite{Guerlin}. In such case, we should introduce
the collective ground state $\left\vert G\right\rangle _{i}=\left\vert
g_{1},...,g_{N_{R}}\right\rangle _{i}$, and the collective excitation state $%
\left\vert R\right\rangle _{i}=\sum_{f}^{N_{R}}\left\vert r_{f}\right\rangle
\left\langle g_{f}\right\vert \otimes \left\vert G\right\rangle _{i}/\sqrt{%
N_{R}}$.

Thus, the first term of the Hamiltonian (\ref{H}) becomes%
\begin{equation}
H_{\text{JC}}=\sum_{i}\left[ \omega a_{i}^{\dag }a_{i}+\epsilon \left\vert
R\right\rangle _{i}\left\langle R\right\vert _{i}+g(a^{\dag }\left\vert
G\right\rangle _{i}\left\langle R\right\vert _{i}+\text{H.c.})\right] .
\label{HJC}
\end{equation}%
The second term in the Hamiltonian (\ref{H}) governs the photon hopping
between two adjacent cavities, and is%
\begin{equation}
H_{\text{HOP}}=-t\sum_{i}(a_{i}^{\dag }a_{i+1}+a_{i+1}^{\dag }a_{i}),
\label{Hop}
\end{equation}%
where $t=\kappa \sqrt{F/2\pi }$\ is the photon hopping rate and $F$\ is the
cavity finesse. The third term in the Hamiltonian (\ref{H}) governs the
long-range van der Waals interaction between Cs Rydberg atoms in different
cavities, and is
\begin{equation}
H_{\text{V}}=\frac{1}{2}\sum_{ij}V(i-j)\left\vert R\right\rangle
_{i}\left\langle R\right\vert _{i}\otimes \left\vert R\right\rangle
_{j}\left\langle R\right\vert _{j},  \label{HV}
\end{equation}%
where $V(i-j)=C_{6}/(x_{i}-x_{j})^{6}$, with $C_{6}$ being the van der Waals
coefficient, and $x_{i}$ being the position of the $i$th cavity\cite{MS10}.
The long-range van der Waals interaction can induce a strong correlation
between Cs Rydberg atoms in different cavities. Hereafter, we use the
nearest-neighbor interaction to represent the entire van der Waals
interaction, i.e., $V\equiv V_{1}$, because $V_{2}=V_{1}/2^{6}$, and $%
V_{3}=V_{1}/3^{6}$, $\cdots $. In the last term of the Hamiltonian (\ref{H}%
), the chemical potential $\mu $ is the Lagrange multiplier, and the total
number of polaritons is $N=\sum_{i}n_{i}=\sum\nolimits_{i}(a_{i}^{\dag
}a_{i}+\left\vert R\right\rangle _{i}\left\langle R\right\vert _{i})$.

It should be noted that a dielectric medium placed near dipoles will alter
the spatial distribution of the electromagnetic field. However, for the
parameters of the nanofiber and Cs Rydberg atoms considered here, this
alteration can be regarded as a higher-order small quantity, compared with
the direct atom-atom interaction\cite{NHA,Mary,Anton}. This allows us to
safely treat the interaction between Cs Rydberg atoms in different cavities
as the standard long-range van der Waals force.\newline

\subsection{Typical parameters.}

Before proceeding, we estimate the relevant parameters of the Hamiltonian (%
\ref{H}) in terms of the above proposal.

\begin{itemize}
\item The effective photon frequency $\omega =\omega _{c}-\omega _{l}$ and
the effective atom transition frequency $\epsilon =E_{r}-E_{g}-\omega
_{l}+\Omega ^{2}/\Delta _{p}$. These two parameters can be well controlled
by the driving frequency $\omega _{l}$ of the classical laser. Thus, these
can have suitable values as required experimentally.

\item The collective atom-photon interaction strength $g=\sqrt{N_{R}}%
g_{0}\Omega /\Delta _{p}$. In our considered nanofiber photonic crystal
cavity, $g_{0}=\sqrt{\eta _{c}\gamma c/L}$, where $\eta _{c}$ is the
channeling efficiency, $c$ is the light velocity, $L$ is the cavity length%
\cite{Fam09,FamL09}. It should be noted that since the Cs Rydberg atoms
considered here are tightly trapped, the decay $\gamma $ of the Rydberg
superatom\ is enhanced\cite{Mandel} by $\gamma =N_{R}\Gamma $, where $\Gamma
$\ is the decay of an isolated Cs Rydberg atom in the state $70S_{1/2}$, due
to the supperradiant effect\cite{Dicke54}. The Rabi frequency and
the detuning are chosen here as $\Omega /2\pi \sim 100$ MHz and $\Delta
_{p}/2\pi \sim 1$ GHz, respectively, which fulfill the adiabatic elimination
condition, $\Delta _{p}\gg \{g_{0}$, $\Omega \}$. In addition, for the
two-color optical dipole trap, with wavelengths\cite{Vetsch} $1064$\ nm and $%
780$\ nm, respectively, the number of Cs Rydberg atoms of each
ensemble can be of the order of $10^{4}$. Therefore, the collective
atom-photon interaction strength reaches $g/2\pi \simeq 2.03$\ GHz, when $%
\eta _{c}/2\pi =0.01$ (see Ref.[33]), $\gamma =27.5$\ MHz ($\Gamma
/2\pi =0.55$\ kHz), $L=10$\ mm, and $N_{R}=$\ $5\times 10^{4}$. If the
atomic number density is increased, this collective atom-photon interaction
strength $g$ can increase rapidly, because it is proportional to $\sqrt{N_{R}%
}$.

\item The van der Waals interaction strength $V(i-j)=C_{6}/(x_{i}-x_{j})^{6}$%
. Based on the aforementioned energy level structures\cite{Singer,Rait}, the
van der Waals coefficient is $C_{6}\approx 610$\ GHz$\cdot \mu $m$^{6}$. For
the distance $x_{i+1}-x_{i}\approx 2.4$\ $\mu $m, the interaction strength
between the nearest-neighbor sites is $V_{1}/2\pi \approx 500$\ MHz, i.e., $%
V/2\pi =V_{1}/2\pi \approx 500$\ MHz. This interaction strength can be
modified by changing the distance of the nearest-neighbor cavities.

\item The cavity decay rate $\kappa $\ and the photon hopping rate $t$. In
the nanofiber photonic crystal cavity considered in Fig.~\ref{fig2}(a)\cite%
{Fam09,FamL09}, $\kappa $ $=\pi c/FL$. In current experimental setups\cite%
{Nayak14}, $F\approx 500$. Thus, $\kappa /2\pi =30$ MHz and $t/2\pi =628$
MHz, when $L=10$ mm. Both the cavity decay rate and the photon hopping rate
can be controlled by changing the cavity length.
\end{itemize}

The above parameters show two basic features: $\left\{ \kappa ,\gamma
\right\} \ll g$ and $V=V_{1}\sim g$. The condition $\left\{ \kappa ,\gamma
\right\} \ll g$ implies that we may safely neglect the influence of the
decay of both cavity and atom, because these only change slightly the phase
boundaries\cite{YWZ13,JR14}. In addition, using the above parameters, we
also estimate that the atomic number density of each cavity is of the order
of $10^{12}$ cm$^{-3}$. For such a typical density, the dephasing time of
the collective states $\left\vert G\right\rangle _{i}$ and $\left\vert
R\right\rangle _{i}$, which are induced by the atomic collision, can, at
least, reach the order of microseconds. This is much larger than the time
scales of $\kappa ^{-1}$\ and $g^{-1}$, and can thus be neglected\cite%
{Rait,Rolf}. This guarantees the validity of our effective two-level model
in Eq.~(\ref{HJC})\cite{Guerlin,Rolf}.

\subsection{Photon long-range repulsive interaction.}

We now construct a strong PLRRI in terms of the Hamiltonian $H_{\text{V}}$.
We begin to address the simplest case, $\kappa =V=0$, in which the
Hamiltonian (\ref{H1}) reduces to
\begin{equation}
H_{\text{S}}=H_{\text{JC}}-\mu N.
\end{equation}%
The eigenstates of the Hamiltonian $H_{\text{S}}$ are given by
\begin{equation}
\left\vert 0-\right\rangle _{i}\equiv \left\vert 0,G\right\rangle _{i}
\label{E0}
\end{equation}%
for $n=0$, and
\begin{equation}
\left\{
\begin{array}{c}
\left\vert n+\right\rangle _{i}=\sin \theta _{n}\left\vert n,G\right\rangle
_{i}+\cos \theta _{n}\left\vert n-1,R\right\rangle _{i} \\
\left\vert n-\right\rangle _{i}=\cos \theta _{n}\left\vert n,G\right\rangle
_{i}-\sin \theta _{n}\left\vert n-1,R\right\rangle _{i}%
\end{array}%
\right.  \label{En}
\end{equation}%
for $n\geqslant 1$, where $\theta _{n}=\arctan (2g\sqrt{n}/\delta )/2$ and $%
\delta =\omega -\epsilon $ is the detuning. The corresponding eigenvalues
are $E_{0}=0$ and
\begin{equation}
E_{n\pm }^{\mu }=n\left( \omega -\mu \right) +\frac{\delta }{2}\pm \left(
\frac{\delta ^{2}}{4}+ng^{2}\right) ^{\frac{1}{2}}\text{ (}n\geqslant 1\text{%
).}  \label{EVA}
\end{equation}%
Since here we investigate the lower-energy behavior, only the lower
polariton branch $\left\vert n-\right\rangle $ is considered\cite{MJH08}.
Thus, the Hamiltonian $H_{\text{S}}$ is rewritten as
\begin{equation}
H_{\text{S}}=\sum_{i}\sum_{n}\left[ n(\omega -\mu )+\frac{\delta }{2}\right]
\left\vert \tilde{n}\right\rangle _{i}\left\langle \tilde{n}\right\vert _{i}%
\mathit{\ }-\sum_{i}\sum_{n}\left( \frac{\delta ^{2}}{4}+ng^{2}\right) ^{%
\frac{1}{2}}\left\vert \tilde{n}\right\rangle _{i}\left\langle \tilde{n}%
\right\vert _{i},  \label{HS}
\end{equation}%
where $\left\vert \tilde{n}\right\rangle _{i}=\left\vert n-\right\rangle
_{i} $. The second term of the Hamiltonian $H_{\text{S}}$ leads to an even
distribution of polaritons, which provides an effective on-site repulsive
interaction between photons\cite{MJH08}. When $t\ll g$, the rotating-wave
approximation is reasonable, and thus the hopping term becomes
\begin{equation}
H_{\text{HOP}}=-t\sum_{n}\sum_{i}\beta _{n,m}\left( \left\vert \tilde{m}%
\right\rangle _{i}\left\langle \tilde{n}\right\vert _{i}\otimes \left\vert
\tilde{n}\right\rangle _{i+1}\left\langle \tilde{m}\right\vert _{i+1}+\text{%
H.c.}\right) ,  \label{HOP}
\end{equation}%
where $\beta _{n,m}=\left( \sqrt{m}\cos \theta _{n}\cos \theta _{m}+\sqrt{n}%
\sin \theta _{n}\sin \theta _{m}\right) ^{2}$\ and $\left\vert \tilde{m}%
\right\rangle _{i}=\left\vert m-\right\rangle _{i}$, with $m=n+1$. In
addition, since the upper polariton branch $\left\vert n+\right\rangle $\
has the higher probability of Rydberg excitation (stronger repulsive
interaction), we also only consider the projection of the van der Waals
interaction into the lower polariton branch $\left\vert n-\right\rangle $.
Thus, the corresponding Hamiltonian becomes
\begin{equation}
H_{\text{V}}^{n,n^{\prime }}=\frac{1}{2}\sum_{ij}\sum_{n,n^{\prime
}>0}J_{n,n^{\prime }}\left( i-j\right) \left\vert \tilde{n}\right\rangle
_{i}\left\langle \tilde{n}\right\vert _{i}\otimes \left\vert \tilde{n}%
^{\prime }\right\rangle _{j}\left\langle \tilde{n}^{\prime }\right\vert _{j}%
\mathbf{,\ }  \label{PLRRI}
\end{equation}%
where
\begin{equation}
J_{n,n^{\prime }}\left( i-j\right) =V\left( i-j\right) \left\langle \tilde{n}%
\right\vert _{i}\left\langle \tilde{n}^{\prime }\right\vert _{j}(\left\vert
R\right\rangle _{i}\left\langle R\right\vert _{i}\otimes \left\vert
R\right\rangle _{j}\left\langle R\right\vert _{j})\left\vert \tilde{n}%
^{\prime }\right\rangle _{j}\left\vert \tilde{n}\right\rangle _{i}=V\left(
i-j\right) \sin ^{2}\theta _{n}\sin ^{2}\theta _{n^{\prime }}  \label{Jij}
\end{equation}%
is the effective interaction strength. Since $V\left( i-j\right) >0$, and
moreover, $V=V_{1}\sim g$, Eq.~(\ref{PLRRI}) demonstrates explicitly that
the van der Waals interaction generates a strong PLRRI. As will be shown
below, this strong PLRRI leads to non-trivial quantum phases exhibiting
photon solid states.\newline

\subsection{Quantum phases.}

We investigate quantum phases and phase diagrams by perturbation theory and
a mapping into an effective Hamiltonian. For instance, when the chemical
potential $\mu $ is weak, the high-occupancy-photon states ($n>1$) of the
Hamiltonian (\ref{H1}) are not considered. In such case, we rewrite the
Hamiltonian (\ref{H1}) in a reduced Hilbert space, with $n=0,1$, as
\begin{eqnarray}
H_{\text{eff}} &=&-J_{\perp }\sum_{i}\left( \left\vert \tilde{1}%
\right\rangle _{i}\left\langle 0\right\vert _{i}\otimes \left\vert
0\right\rangle _{i+1}\left\langle \tilde{1}\right\vert _{i+1}+\text{H.c.}%
\right) +\frac{1}{2}\sum_{ij}J_{\parallel }\left( i-j\right) \left\vert
\tilde{1}\right\rangle _{i}\left\langle \tilde{1}\right\vert _{i}\otimes
\left\vert \tilde{1}\right\rangle _{j}\left\langle \tilde{1}\right\vert _{j}
\notag \\
&&+E_{1-}^{\mu }\sum_{i}\left\vert \tilde{1}\right\rangle _{i}\left\langle
\tilde{1}\right\vert _{i},  \label{Heff}
\end{eqnarray}%
where $J_{\perp }=t\cos ^{2}\theta _{1}$, $J_{\parallel }\left( i-j\right)
=J_{1,1}\left( i-j\right) $, and $E_{1-}^{\mu }=\omega -\mu +\delta /2-\sqrt{%
(\delta /2)^{2}+g^{2}}$ is the single-particle energy of the $\left\vert
\tilde{1}\right\rangle $ state. This effective photon hopping rate $J_{\perp
}$\ can be easily tuned by the detuning $\delta $, since $\theta
_{1}=\arctan (2g/\delta )/2$. In addition, for the low-energy effective
Hamiltonian (\ref{Heff}), it is convenient to introduce a renormalized
nearest-neighbor van der Waals interaction $\tilde{V}=V\sin ^{4}\theta _{1}$%
\ to simplify the discussions about phase diagrams, as shown below.

\begin{figure}[t]
\centering\includegraphics[width = 1.0\linewidth]{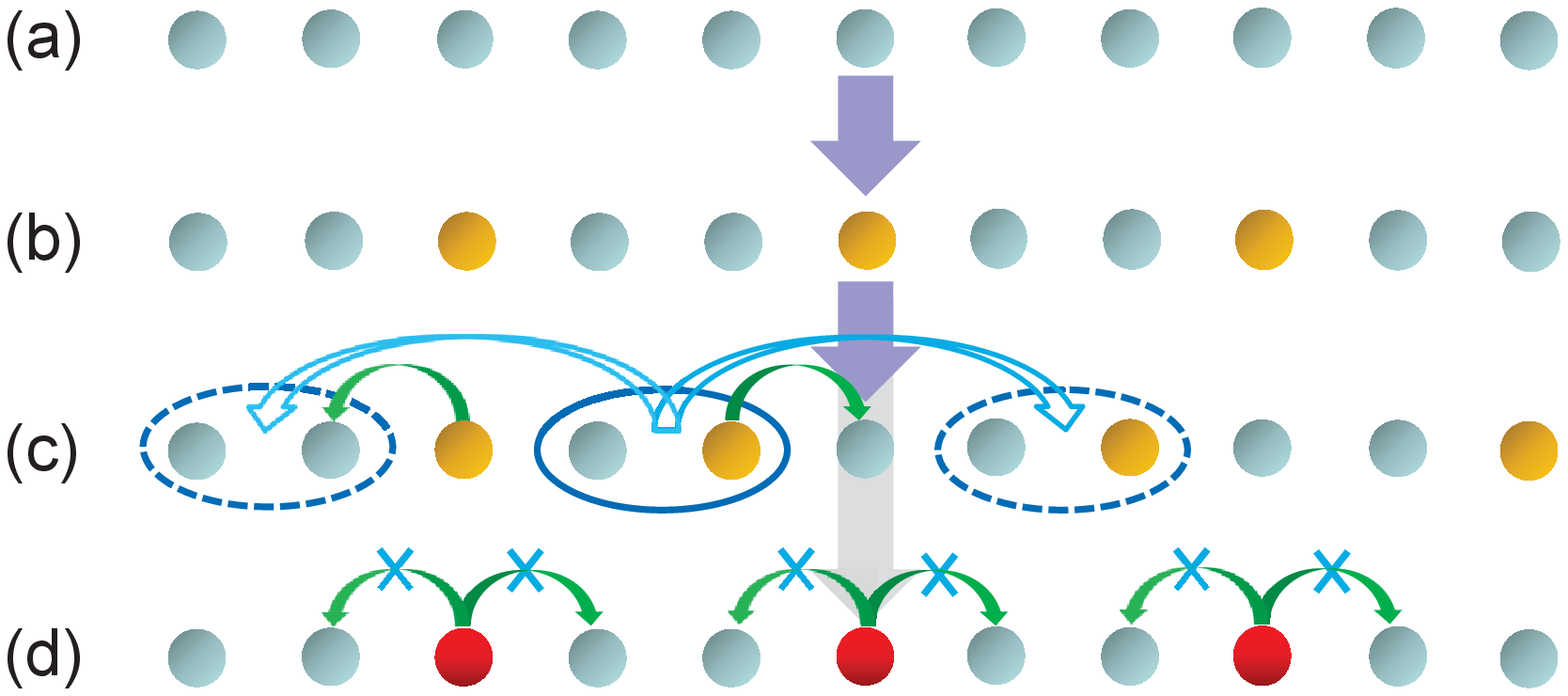}
\caption{\textbf{Photon distributions of each cavity for different effective
strengths $V$ of the van der Waals interaction, when increasing the chemical
potential $\protect\mu $.}}
\label{fig3}
\end{figure}

We first consider the case without photon hopping ($J_{\perp }=0$). At the
initial time, we assume that every cavity is in its vacuum state, as shown
in Fig.~\ref{fig3}(a). When increasing the chemical potential $\mu $,
photons in some cavities can be excited, due to the existence of the PLRRI
(without the PLRRI, all cavities are excited identically\cite{MJH08}), and
some $\left\vert \tilde{1}\right\rangle $ states emerges, as shown in Fig.~%
\ref{fig3}(b). The corresponding critical point is
\begin{equation}
\frac{\mu _{c0}-\omega }{g}=\frac{\delta }{2g}-\left( 1+\frac{\delta ^{2}}{%
4g^{2}}\right) ^{\frac{1}{2}},  \label{EUc}
\end{equation}%
derived from $E_{1-}^{\mu }(\mu _{c0})=0$. Since the $\left\vert \tilde{1}%
\right\rangle $ states are generated one by one and deviate from each other,
the system exhibits photon solid states, which are mainly governed by
different filling factors
\begin{equation}
\rho =\frac{p}{q}\text{ \ \ \ }(\leq 1),  \label{Rou}
\end{equation}%
with $p$ and $q$ being both integers. In order to quantitatively determine
the filling factor $\rho $, we introduce $X_{i}^{0}$ and $X_{i}^{l}$, where $%
X_{i}^{0}$ is the position of the $i$th $\left\vert \tilde{1}\right\rangle $
state and $X_{i}^{l}$ is the distance to the $l$th next $\left\vert \tilde{1}%
\right\rangle $ state, satisfying $X_{i}^{l}$ $=X_{i+l}^{0}-X_{i}^{0}$. When
the ground-state energy is minimized for all sites, we have
\begin{equation}
X_{i}^{l}=r_{l}\text{ \ \ \ or \ \ \ }r_{l}+1,  \label{R1}
\end{equation}%
where $r_{l}<l/\rho <r_{l}+1$, and satisfy the relation\cite%
{Hubbard,Pokrovsky}
\begin{equation}
\sum_{i}X_{i}^{l}=lN_{0}.  \label{R2}
\end{equation}%
In Eq.~(\ref{R2}), $N_{0}$\ is the total number of cavities. For a given
filling state, the repulsive interaction energy of the $\left\vert \tilde{1}%
\right\rangle $ states can be estimated by applying the relations in Eqs.~(%
\ref{R1})-(\ref{R2}) to the Hamiltonian (\ref{Heff}). Moreover, the
corresponding phases are stable if it costs energy to add or remove a
particle and rearrange the structure.

\subsection{Photon solid phase.}

We define the photon solid phase, with the filling factor $\rho $, as $%
\left\vert c\right\rangle _{q}$. If we add one $\left\vert \tilde{1}%
\right\rangle $ state, $\left\vert c\right\rangle _{q}$ becomes $\left\vert
p\right\rangle _{q}$ and the $\left\vert \tilde{1}\right\rangle $\ states
are crowded. To minimize the repulsive energy, the summation of distances
between the $\left\vert \tilde{1}\right\rangle $\ states must be a minimum.
Thus, the most likely rearrangement structure is that some pairs of the
adjacent $\left\vert \tilde{1}\right\rangle $\ states are shortened by one
site\cite{Hubbard,Bak}. By considering the periodic boundary condition and
relations in Eqs.~(\ref{R1})-(\ref{R2}), $r_{l}$ $\left\vert \tilde{1}%
\right\rangle $ state pairs with $X_{i}^{l}=$ $(r_{l}+1)$ must be replaced
by $(r_{l}+1)$ $\left\vert \tilde{1}\right\rangle $ state pairs with $%
X_{i}^{l}=r_{l}$. In addition, at the phase-transition point, there is no
energy gap\cite{Bak} between $\left\vert c\right\rangle _{q}$ and $%
\left\vert p\right\rangle _{q}$, i.e., $E(\left\vert c\right\rangle
_{q})=E(\left\vert p\right\rangle _{q})$, and the critical point is thus
obtained by
\begin{eqnarray}
\mu _{\rho }^{0}\left( p\right) &=&\omega +\frac{\delta }{2}-\left( \frac{%
\delta ^{2}}{4}+g^{2}\right) ^{\frac{1}{2}}+\sum_{k=1,k\neq fp}\left[ \left(
r_{k}+1\right) J_{\parallel }(r_{k})-r_{k}J_{\parallel }(r_{k}+1)\right] +
\label{u1} \\
&&\sum_{k=1}\left[ kqJ_{\parallel }(kq-1)-\left( kq-1\right) J_{\parallel
}(kq)\right] ,  \notag
\end{eqnarray}%
where $f$\ is any integer (see Methods section). Similarly, if we remove one
$\left\vert \tilde{1}\right\rangle $ state, $\left\vert c\right\rangle _{q}$
turns into $\left\vert h\right\rangle _{q}$, and the corresponding critical
point is given by (see Methods section)
\begin{eqnarray}
\mu _{\rho }^{0}\left( h\right) &=&\omega +\frac{\delta }{2}-\left( \frac{%
\delta ^{2}}{4}+g^{2}\right) ^{\frac{1}{2}}+\sum_{k=1,k\neq fp}\left[ \left(
r_{k}+1\right) J_{\parallel }(r_{k})-r_{k}J_{\parallel }(r_{k}+1)\right] +
\label{u2} \\
&&\sum_{k=1}\left[ \left( kq+1\right) J_{\parallel }(kq)-kqJ_{\parallel
}(kq+1)\right] .  \notag
\end{eqnarray}%
In terms of the obtained $\mu _{\rho }^{0}\left( p\right) $ and $\mu _{\rho
}^{0}\left( h\right) $, the stability interval, $\Delta \mu _{\rho }=\mu
_{\rho }^{0}\left( p\right) -\mu _{\rho }^{0}\left( h\right) $, is evaluated
as
\begin{equation}
\Delta \mu _{\rho }=\sum_{k=1}kqJ_{\parallel }(kq+1)+kqJ_{\parallel
}(kq-1)-2kqJ_{\parallel }(kq).  \label{du}
\end{equation}%
The expression for $\Delta \mu _{\rho }$ shows that the stability interval
is only dependent on $q$, and moreover, decreases rapidly when increasing $q$%
. This means that the photon solid phases with $p=1$, i.e., $\rho =1/q=1/2$,
$1/3$, $1/4$,$\cdots $, are more likely to be observed. Below, we mainly
address these phases.\newline

\subsection{Photon Devil's staircase.}

\begin{figure}[t]
\centering\includegraphics[width = 1.0\linewidth]{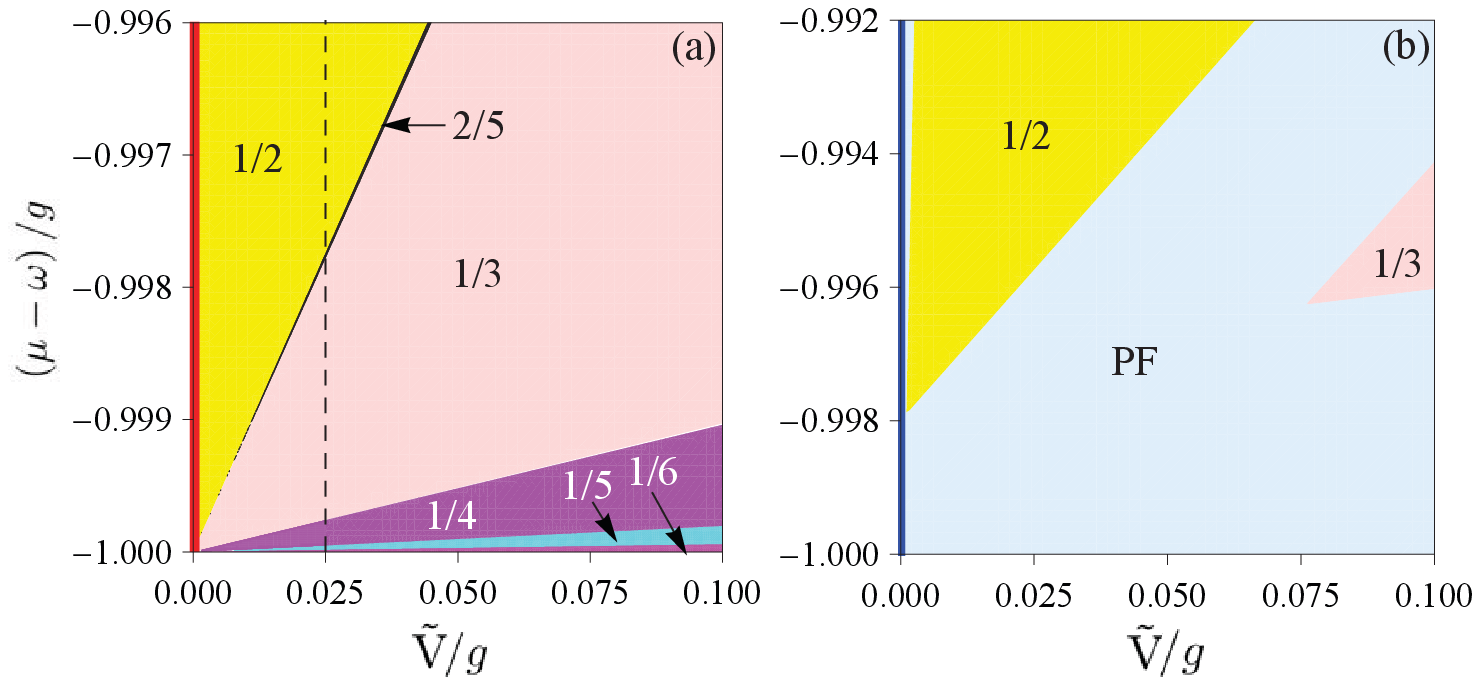}
\caption{\textbf{The filling factor $\protect\rho =p/q$ as a function of the
chemical potential $\protect\mu $\ and the renormalized effective strength }$%
\mathbf{\tilde{V}}=\mathbf{V}\sin ^{4}\protect\theta _{1}$\textbf{\ of the
van der Waals interaction, when} \textbf{(a) }$J_{\bot }/g=0$\textbf{\ and
(b) }$J_{\bot }/g=0.001$.}
\label{fig4}
\end{figure}

In Fig.~\ref{fig4}(a), we plot the filling factor $\rho $ as a function of
the chemical potential $\mu $\ and the renormalized effective strength $%
\tilde{V}=V\sin ^{4}\theta _{1}$ of the van der Waals interaction, in terms
of the obtained $\mu _{\rho }^{0}\left( p\right) $ and $\mu _{\rho
}^{0}\left( h\right) $ in Eqs.~(\ref{u1}) and~(\ref{u2}). For $\tilde{V}=0$,
$\rho =1$, as expected [see the red solid line in Fig.~\ref{fig4}(a)].
However, the results for finite $\tilde{V}$\ [for example, $\tilde{V}=0.025g$%
; see the black dashed line in Fig.~\ref{fig4}(a)] are very interesting.
When increasing $\mu $, $\rho $ is not a constant, but varies
\textquotedblleft jumpily\textquotedblright\ from $1/6$, $1/5$, $1/2$, $1/4$%
, $1/3$, $2/5$, to $1/2$. The reason is that when increasing $\mu $, $%
E_{1-}^{\mu }$ decreases, and excitation of the cavities is thus favorable.
This behavior clearly shows a Devil's staircase\cite{Bak,CR}. Moreover, this
Devil's staircase could be detected experimentally by measuring the
mean-photon number $\left\langle a^{\dag }a\right\rangle /L$, since $%
\left\langle a^{\dag }a\right\rangle /L=\rho /2$, and thus here called the
\textit{photon Devil's staircase}. However, when increasing $\tilde{V}$, $%
\rho $\ varies jumpily from high to low because the PLRRI prevents the
photon excitation.

Recently, the photon nearest-neighbor interaction was studied and a photon
solid state was predicted\cite{JJ13}. In that case, the $Z_{2}$\ symmetry,
translated by one site, has been broken. Here the PLRRL generates a
long-range translation symmetry, whose breaking induces the photon Devil's
staircase. Moreover, it leads to other non-trivial phases when the photon
hopping exists.

Notice that between the adjacent photon solid phases, with $\rho =1/q$\ and $%
\rho =1/(q\mp 1)$, respectively, there are many transition states which have
different numbers of defects. Here we define the pairs of the $\left\vert
\tilde{1}\right\rangle $\ states with shorter (longer) distance as a
particle- (hole-) like defect structure. Since these states have very small
stability intervals, they should be hard to observe when $J_{\perp }=0$, and
thus not plotted in Fig.~\ref{fig3}(b). However, when $J_{\perp }\neq 0$,
they play an important role for the ground-state properties, because of the
motion of the defects, as shown in Fig.~\ref{fig3}(c). Especially, when the
hopping energy is negative, the states with defects may be more stable than
the adjacent photon solid states. Thus, the photon solid phases melt and a
photon-floating solid phase\cite{Fendley} can emerge. In general, it is
difficult to fully characterize this process. However, in the region close
to the phase-transition point, the repulsive interaction between the defects
only allow one defect. Thus, the phase boundary can be estimated by
comparing the energy of the photon solid state $\left\vert c\right\rangle
_{q}$\ with that of the state with one defect. Using a perturbative method,
we obtain the following phase boundaries (see Methods section):
\begin{equation}
\mu _{\rho }^{\text{up}}=\mu _{\rho }^{0}\left( p\right) -2qJ_{\perp }\text{%
, \ }\mu _{\rho }^{\text{down}}=\mu _{\rho }^{0}\left( h\right) +2qJ_{\perp
}.  \label{B1}
\end{equation}

Equation (\ref{B1}) shows that the hopping energies of the defects reduce
the regions where the photon solid phases exist, because $\mu _{\rho }^{%
\text{up}}$\ $-\mu _{\rho }^{\text{down }}=\Delta \mu _{\rho }-4qJ_{\perp }$%
. In particular, when $q\geqslant \Delta \mu _{\rho }/\left( 4J_{\perp
}\right) $, $\mu _{\rho }^{\text{up}}$\ $\leqslant \mu _{\rho }^{\text{down}%
} $, and thus the energy bands of the particle- and hole-like defect states
cross and the photon solid phases cannot exist. This is the reason why only
the photon solid phases, with $\rho =1/2$\ and $\rho =1/3$, can emerge in
Fig.~\ref{fig4}(b). From Fig.~\ref{fig4}(b), we also see that the regions
where the photon solid phases exist are very small, and are melted for a
smaller $J_{\perp }$ ($J_{\perp }/g=0.001$). This implies that the hopping
term can be treated as a perturbation. So the results from the phase
boundaries in Eq.~(\ref{B1}) are reasonable. Strictly speaking, in the
photon-floating solid phase, the total number of the $\left\vert \tilde{1}%
\right\rangle $ states is sensitive to the fluctuation of the parameters,
and also $\rho $ and $\left\langle a^{\dag }a\right\rangle /L$ are hard to
calculate in that phase. Recently, the quantum Monte Carlo method has been
used to solve this problem\cite{JZ08}. When $\tilde{V}=0$, the
photon-floating solid phase disappears [see the blue line in Fig.~\ref{fig4}%
(b)].\newline

\subsection{Photon-frozen solid phase.}

\begin{figure}[t]
\centering\includegraphics[width = 1.0\linewidth]{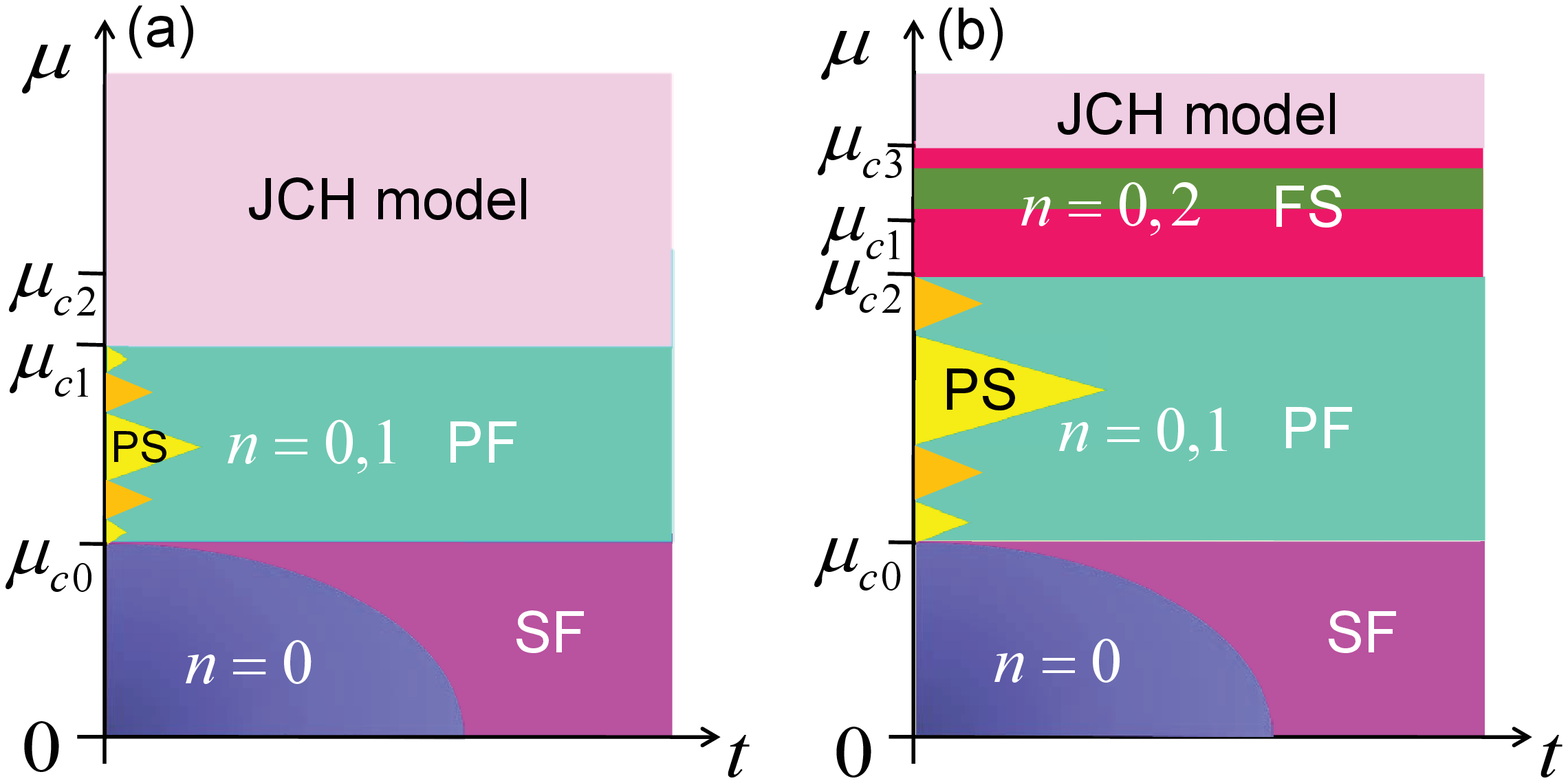}
\caption{\textbf{Schematics of the ground-state phase diagrams as functions
of the chemical potential $\protect\mu $ and the photon hopping rate }$t$%
\textbf{, when }$\protect\delta =0$.}
\label{fig5}
\end{figure}

Finally, we address the case of a strong chemical potential $\mu $, in which
the higher-photon-occupancy states in some cavities can occur, and moreover,
the single-particle energy of the $\left\vert \tilde{2}\right\rangle $
state, $E_{2-}^{\mu }$, is close to that of the $\left\vert \tilde{1}%
\right\rangle $ state, $E_{1-}^{\mu }$, (here we omit the case $n>2$). In
this case, there are three kinds of repulsive interactions: between the $%
\left\vert \tilde{1}\right\rangle $\ and $\left\vert \tilde{1}\right\rangle $%
\ states, between the $\left\vert \tilde{2}\right\rangle $\ and $\left\vert
\tilde{2}\right\rangle $\ states, and between the $\left\vert \tilde{1}%
\right\rangle $\ and $\left\vert \tilde{2}\right\rangle $\ states. Moreover,
the photon hopping has two channels, from the $\left\vert 0\right\rangle $\
to $\left\vert \tilde{1}\right\rangle $\ states and from the $\left\vert
\tilde{1}\right\rangle $\ to $\left\vert \tilde{2}\right\rangle $\ states.
These two channels are very complex. However, in the resonant case ($\delta
=0$), $\sin ^{2}\theta _{n}=1/2$, and $H_{\text{V}}^{n,n^{\prime }}$ is thus
independent of $n$. This indicates that the photon numbers of the excited
cavities are only determined by $E_{1-}^{\mu }$ and $E_{2-}^{\mu }$. When
the PLRRI is not sufficiently strong, the lattice can be fully filled in the
weak-$\mu $\ region. In this region, $E_{2-}^{\mu }>E_{1-}^{\mu }$, and the
ground state, still governed by the Hamiltonian (\ref{Heff}), is thus
composed of the $\left\vert 0\right\rangle $ and $\left\vert \tilde{1}%
\right\rangle $ states. By increasing $\mu $, $\rho $ increases from $0$ and
reaches $1$. Further increasing $\mu $, all cavities can be excited with
uniform photon numbers, which is similar to that of the standard
Jaynes-Cummings-Hubbard model, as shown in Fig.~\ref{fig5}(a).

However, there is a non-trivial case for a strong PLRRI, as shown in Fig.~%
\ref{fig5}(b). In such case, the photon solid phases can exist in the strong-%
$\mu $\ region. But we cannot ensure that the lattice is fully filled by the
$\left\vert \tilde{1}\right\rangle $ states, due to inversion of $%
E_{1-}^{\mu }$ and $E_{2-}^{\mu }$. This process can be determined by
comparing $\mu _{c1}\approx \omega -g+1.0175V$, obtained by making $\rho =1$%
\ in $\mu _{\rho }^{0}\left( h\right) $, with the other critical point $\mu
_{c2}\approx \omega +0.414g$ (the degenerate point of $E_{1-}^{\mu _{c2}}$
and $E_{2-}^{\mu _{c2}}$). When $V>0.576g$, $\mu _{c1}>\mu _{c2}$, and there
is a transition from the $\left\vert \tilde{1}\right\rangle $\ to $%
\left\vert \tilde{2}\right\rangle $ states in the excited cavities. Thus,
this transition induces a new crystalline configuration, which is composed
of the $\left\vert 0\right\rangle $\ and $\left\vert \tilde{2}\right\rangle $
states. The corresponding low-energy behavior is governed by a new effective
Hamiltonian
\begin{equation}
H_{\text{eff}}^{\prime }=\frac{1}{2}\sum_{ij}J_{\parallel }\left( i-j\right)
\left\vert \tilde{2}\right\rangle _{i}\left\langle \tilde{2}\right\vert
_{i}\otimes \left\vert \tilde{2}\right\rangle _{j}\left\langle \tilde{2}%
\right\vert _{j}+E_{2-}^{\mu }\sum_{i}\left\vert \tilde{2}\right\rangle
_{i}\left\langle \tilde{2}\right\vert _{i},\   \label{NEH}
\end{equation}%
where $J_{\parallel }\left( i-j\right) =J_{2,2}\left( i-j\right)
=J_{1,1}\left( i-j\right) $,\ and $E_{2-}^{\mu }=2(\omega -\mu )-\sqrt{2}g$.
Since
\begin{equation}
\left\langle 0\right\vert _{i+1}\left\langle \tilde{2}\right\vert
_{i}(a_{i}^{\dag }a_{i+1})\left\vert 0\right\rangle _{i}\left\vert \tilde{2}%
\right\rangle _{i+1}=0,  \label{ii2}
\end{equation}%
the photon hopping is always frozen even if $t$ exists. We denote the
corresponding phase as the \textit{photon-frozen solid phase}. In this
phase, the fractional filling structure of the $\left\vert \tilde{2}%
\right\rangle $ states is robust, i.e., it is not easily destroyed by the
photon hopping. In terms of the Hamiltonian (\ref{NEH}), when further
increasing $\mu $ to satisfy $\mu >\mu _{c3}\approx (2\omega -\sqrt{2}%
g+1.0175V)/2$, the lattice can be fully filled by the $\left\vert \tilde{2}%
\right\rangle $ states, as shown in Fig.~\ref{fig5}(b).\newline

\section*{{\protect\LARGE \textbf{Discussion}}}

In summary, we have achieved a strong PLRRI by controlling the van der Waals
interaction of Rydberg atoms located in different cavities in extended
Jaynes-Cummings-Hubbard lattices, and then predicted novel quantum phases.
Since the atom-cavity polariton can be easily controlled experimentally\cite%
{FB07,MP14}, our proposal offers a new way to control the interaction
between individual photons. In addition, our proposal might help to explore
rich many-body phenomena of light and quantum nonlinear optics, as well as
potential applications to quantum information and computing.\newline

\newpage
\section*{{\protect\LARGE \textbf{Methods}}}

\subsection{Derivation of Eqs.~(\protect\ref{u1}) and~(\protect\ref{u2}).}

We have described the low-energy behavior of the Hamiltonian (\ref{H}) by an
effective Hamiltonian (\ref{Heff}). Moreover, we have also pointed out that
when $J_{\perp }=0$, there is a succession of photon crystal states with
different filling factors, denoted as a photon Devil's staircase structure,
and the energy gap of the photon crystal states can be calculated in terms
of Eqs.~(\ref{R1}) and~(\ref{R2}), i.e., $X_{i}^{l}=r_{l}$ or $r_{l}+1$, and
$\sum_{i}X_{i}^{l}=lN_{0}$. For example, we define the crystalline ground
state, with the filling factor $\rho =p/q$, as $\left\vert c\right\rangle
_{q}$. By adding one $\left\vert \tilde{1}\right\rangle $ state, the
crystalline ground state $\left\vert c\right\rangle _{q}$ becomes $%
\left\vert p\right\rangle _{q}$. After rearranging the $\left\vert \tilde{1}%
\right\rangle $ states, the distance $r_{l}$ between the $\left\vert \tilde{1%
}\right\rangle $ states is changed. Using Eqs.~(\ref{R1}) and~(\ref{R2}), $%
r_{l}$ $\left\vert \tilde{1}\right\rangle $ state pairs with $X_{i}^{l}=$ $%
(r_{l}+1)$ must be replaced by $(r_{l}+1)$ $\left\vert \tilde{1}%
\right\rangle $ state pairs with $X_{i}^{l}=r_{l}$. So the corresponding
energy shift, $\Delta E^{+}=E(\left\vert p\right\rangle _{q})-E(\left\vert
c\right\rangle _{q})$, is calculated as
\begin{eqnarray}
\Delta E^{+} &=&E_{1-}^{\mu }+\left( r_{1}+1\right) J_{\parallel
}(r_{1})-r_{1}J_{\parallel }(r_{1}+1)+\left( r_{2}+1\right) J_{\parallel
}(r_{2})-r_{2}J_{\parallel }(r_{2}+1)+\cdots  \label{EUP} \\
&&+qJ_{\parallel }(q-1)-(q-1)J_{\parallel }(q)+\cdots +2qJ_{\parallel
}(2q-1)-(2q-1)J_{\parallel }(2q)+\cdots ,  \notag
\end{eqnarray}%
where $r_{p}=q$, $r_{2p}=2q$,\ldots , have been inserted\cite{Bak}.
Similarly, by removing one $\left\vert \tilde{1}\right\rangle $ state from $%
\left\vert c\right\rangle _{q}$, we obtain a new state $\left\vert
h\right\rangle _{q}$. The corresponding energy shift, $\Delta
E^{-}=E(\left\vert h\right\rangle _{q})-E(\left\vert c\right\rangle _{q})$,
is calculated as
\begin{eqnarray}
\Delta E^{-} &=&-E_{1-}^{\mu }-\left( r_{1}+1\right) J_{\parallel
}(r_{1})+r_{1}J_{\parallel }(r_{1}+1)-\left( r_{2}+1\right) J_{\parallel
}(r_{2})+r_{2}J_{\parallel }(r_{2}+1)+\cdots  \label{EL} \\
&&-(q+1)J_{\parallel }(q)+qJ_{\parallel }(q+1)-\cdots -(2q+1)J_{\parallel
}(2q)+2qJ_{\parallel }(2q+1)+\cdots .  \notag
\end{eqnarray}%
These equations govern the energy gap of the photon crystal state $%
\left\vert c\right\rangle _{q}$. Obviously, at the phase-transition point,
the energy gap is closed, i.e., $\Delta E^{\pm }=0$. Using the expression $%
E_{1-}^{\mu }=\left( \omega -\mu \right) +\delta /2-\sqrt{\delta ^{2}/4+g^{2}%
}$, we can derive the critical point of the chemical potential. The critical
point between $\left\vert c\right\rangle _{q}$ and $\left\vert
p\right\rangle _{q}$ is%
\begin{eqnarray}
\mu _{\rho }^{0}\left( p\right) &=&\omega +\frac{\delta }{2}-\left( \frac{%
\delta ^{2}}{4}+g^{2}\right) ^{\frac{1}{2}}  \label{CUP} \\
&&+\sum_{k=1,k\neq fp}\left[ \left( r_{k}+1\right) J_{\parallel
}(r_{k})-r_{k}J_{\parallel }(r_{k}+1)\right] +\sum_{k=1}\left[
kqJ_{\parallel }(kq-1)-\left( kq-1\right) J_{\parallel }(kq)\right] ,  \notag
\end{eqnarray}%
where $f$ is any integer. Similarly, the critical point between $\left\vert
c\right\rangle _{q}$ and $\left\vert h\right\rangle _{q}$ is given by%
\begin{eqnarray}
\mu _{\rho }^{0}\left( h\right) &=&\omega +\frac{\delta }{2}-\left( \frac{%
\delta ^{2}}{4}+g^{2}\right) ^{\frac{1}{2}}  \label{CDO} \\
&&+\sum_{k=1,k\neq fp}\left[ \left( r_{k}+1\right) J_{\parallel
}(r_{k})-r_{k}J_{\parallel }(r_{k}+1)\right] +\sum_{k=1}\left[ \left(
kq+1\right) J_{\parallel }(kq)-kqJ_{\parallel }(kq+1)\right] .  \notag
\end{eqnarray}%
\newline

\subsection{Derivation of Eq.~(5).}

We define
\begin{equation}
\left\vert \tilde{p}\right\rangle _{q}=\sum_{i=1}^{L/q}C_{i}\left\vert
p\right\rangle _{q}^{i}\   \label{State}
\end{equation}%
as a state with a one particle-like defect, where the index $i$\ denotes the
position of the defect and $C_{i}$\ is its coefficient. For simplicity, we
only consider the lowest order of the photon hopping: the motion of the
defect. Inserting $\left\vert \tilde{p}\right\rangle _{q}$ into equation $%
E(\left\vert \tilde{p}\right\rangle _{q})=\left\langle \tilde{p}\right\vert
_{q}H_{\text{eff}}\left\vert \tilde{p}\right\rangle _{q}$, we obtain%
\begin{equation}
E(\left\vert \tilde{p}\right\rangle _{q})=E^{0}(\left\vert \tilde{p}%
\right\rangle _{q})-2qJ_{\perp }\cos (\tilde{k}q),  \label{Ep}
\end{equation}%
where $E^{0}(\left\vert \tilde{p}\right\rangle _{q})$\ is the summation of
the on-site and repulsive energies, $-2qJ_{\perp }\cos (\tilde{k}q)$\ is the
hopping energy band of a defect with wave number $\tilde{k}$. The phase
boundary is determined by the lowest energy of $\left\vert \tilde{p}%
\right\rangle _{q}$, i.e., $\tilde{k}=0$ and\textbf{\ }$%
E(\left\vert c\right\rangle _{q})=E^{0}(\left\vert \tilde{p}\right\rangle
_{q})-2qJ_{\perp }$. Thus, the upper bounds of the photon solid phases are
given by
\begin{equation}
\mu _{\rho }^{\text{up}}=\mu _{\rho }^{0}\left( p\right) -2qJ_{\perp }.
\label{up}
\end{equation}%
Similar to the above discussions, the lower bounds of the photon solid
phases are obtained by
\begin{equation}
\mu _{\rho }^{\text{down}}=\mu _{\rho }^{0}\left( h\right) +2qJ_{\perp }.
\label{down}
\end{equation}

\begin{addendum}

 \item This work is supported in part by the 973
program under Grant No.~2012CB921603; the NNSFC under Grant No.~11422433, No.~11434007, No.~61275211, and No.~91436108; the PCSIRT under Grant No.~IRT13076; the NCET under Grant No.~13-0882; the FANEDD under Grant No.~201316;
the OIT under Grant No.~2013804; OYTPSP; and SSCC. FN is partially supported
by the RIKEN iTHES Project, the MURI Center for Dynamic Magneto-Optics via the AFOSR award number FA9550-14-1-0040, the Impact program of JST, and a
Grant-in-Aid for Scientific Research (S).

 \item[Author Contributions] Y.Z., J.F., J.Q.L., J.M., G.C., S.J. and F.N. conceived the idea, Y.Z. and J.F. performed the calculation, G.C., S.J. and F.N. wrote the manuscript. Y.Z. and J.F. contributed equally to this work.

 \item[Competing Interests] The authors declare that they have no
competing financial interests.

\end{addendum}
\newpage

\textbf{Figure 1: Schematic diagram of the system studied.} A 1D nanofiber
photonic crystal cavity array, with an ensemble of Cs Rydberg atoms (red
disks) placed near each nanofiber. Photons can hop between two adjacent
cavities, indicated by green double-arrows. FBG denotes the fiber Bragg
grating.

\textbf{Figure 2:} \textbf{(a) The sectional plot of the \emph{i}th
atom-cavity interaction system, and (b) energy levels of a single
three-level Cs Rydberg atom and their transition.} In (a), the yellow and
green solid curves schematically show the intensity distributions of the
intracavity and evanescent fields, respectively. $b$ denotes the radius of
the nanofiber, which is about $0.25$ $\mu $m, and $L$ is the length of
cavity. In general, the radius $b$ is smaller than the distance of the
nearest-neighbor cavities, which is chosen here as $x_{i+1}-x_{i}\approx 2.4$%
\ $\mu $m. In addition, FBG denotes the fiber Bragg grating. In (b), the
green-arrowed line shows the photon-induced transition, whereas the
red-arrowed line labels the other transition governed by the classical
driving laser. The detunings are given by $\Delta _{p}=(E_{p}-E_{g})-\omega
_{c}$ and $\Delta _{r}=\omega _{l}-(E_{r}-E_{p})$, respectively.

\textbf{Figure 3: Photon distributions of each cavity for different
effective strengths $V$ of the van der Waals interaction, when increasing
the chemical potential $\mu $.} (a-b) $t=0$ with a weak $V$, (c) $t\neq 0$
with a weak $V$, and (d) $t\neq 0$ with a large $\mu $ and a strong $V$. The
vacuum state $\left\vert 0\right\rangle $ is denoted by light blue disks,
and the photon excitation state $\left\vert \tilde{1}\right\rangle $ is
shown in orange. (a) In the initial state, every cavity is in its vacuum
state. When increasing $\mu $, cavities can be excited. Due to existence of
the PLRRI, the $\left\vert \tilde{1}\right\rangle $ states are generated one
by one and deviated from each other. Thus, the ground states of system are a
series of photon solid phases, with different fraction filling factors (from
low to high). We call it photon Devil's stair case. As an example, (b) shows
a photon solid phase with a period of $3$ sites ($\cdots \left\vert
0\right\rangle \left\vert 0\right\rangle \left\vert \tilde{1}\right\rangle
\left\vert 0\right\rangle \left\vert 0\right\rangle \left\vert \tilde{1}%
\right\rangle \cdots $). (c) Melting of this photon solid phase. A
particle-like defect with the unit cell $\left\vert 0\right\rangle
\left\vert \tilde{1}\right\rangle $ is shown inside the blue solid elliptic
curve in (c). When a photon on the edge of the defect hops one site, this
defect will move three sites (the new possible positions are labeled by
dashed ellipses). (d) Plot of a photon-frozen solid phase, which is composed
of the $\left\vert 0\right\rangle $and $\left\vert \tilde{2}\right\rangle $
(red color) states.

\textbf{Figure 4: The filling factor $\rho =p/q$ as a function of the
chemical potential $\mu $\ and the renormalized effective strength }$\mathbf{%
\tilde{V}}=\mathbf{V}\sin ^{4}\theta _{1}$\textbf{\ of the van der Waals
interaction, when} \textbf{(a) }$J_{\bot }/g=0$\textbf{\ and (b) }$J_{\bot
}/g=0.001$. In (a), the ground states of system are the photon solid phases.
For finite $\tilde{V}$, when increasing $\mu $, excitation of the cavities
is favorable, and $\rho $\ varies \textquotedblleft
jumpily\textquotedblright\ from $1/6$, $1/5$, $1/2$, $1/4$, $1/3$, $2/5$, to
$1/2$. This behavior clearly shows a devil's staircase. On the contrary,
when increasing $\tilde{V}$ for a finite $\mu $, the PLRRI prevents
excitation of the cavities, and $\rho $ decreases \textquotedblleft
jumpily\textquotedblright\ from $1/2$ to $1/6$. In (b), when the photon
hopping exists, the photon solid phases melt, attributed to the motion of
particle- and hole-like defects. Thus, the photon-floating solid phase (PF)
emerges.

\textbf{Figure 5: Schematics of the ground-state phase diagrams as functions
of the chemical potential $\mu $ and the photon hopping rate }$t$\textbf{,
when }$\delta =0$\textbf{.} In (a), the PLRRI is weak and all cavities are
excited to the $\left\vert \tilde{1}\right\rangle $\ states before the
higher-photon-occupancy states emerge. This can be determined by considering
$\mu _{c1}<\mu _{c2}$. In (b), the PLRRI is strong and the photon-frozen
solid phase occurs. This can be determined by considering $\mu _{c1}>\mu
_{c2}$. When $\mu >\mu _{c1}$ and $\mu >\mu _{c3}$, all cavities in (a) and
(b) are excited identically, respectively. Here, SF, PS, PF, and FS denote
the following phases: superfluid, photon solid, photon-floating solid, and
photon-frozen solid, respectively. JCH stands for Jaynes-Cummings-Hubbard.
This figure is not to scale.


\newpage

\newpage

\newpage

\newpage

\newpage

\newpage


\end{document}